# Generation of micro-J pulses in the deep UV at MHz repetition rates

F. KÖTTIG,[1,*] F. TANI,[1] J. C. TRAVERS,[1,2] AND P. ST.J. RUSSELL[1]

[1]Max Planck Institute for the Science of Light, Staudtstrasse 2, 91058 Erlangen, Germany
[2]School of Engineering and Physical Sciences, Heriot-Watt University, Edinburgh EH14 4AS, United Kingdom
*Corresponding author: felix.koettig@mpl.mpg.de

**Although ultraviolet (UV) light is important in many areas of science and technology, there are very few if any lasers capable of delivering wavelength-tunable ultrashort UV pulses at MHz repetition rates. Here we report the generation of deep-UV laser pulses at MHz repetition rates and µJ-energies by means of dispersive wave (DW) emission from self-compressed solitons in gas-filled single-ring hollow-core photonic crystal fiber (SR-PCF). Pulses from an ytterbium fiber laser (~300 fs) are first compressed to ~25 fs in a SR-PCF-based nonlinear compression stage, and subsequently used to pump a second SR-PCF stage for broadband DW generation in the deep UV. The UV wavelength is tunable by selecting the gas species and the pressure. At 100 kHz repetition rate, a pulse energy of 1.05 µJ was obtained at 205 nm (average power 0.1 W), and at 1.92 MHz, a pulse energy of 0.54 µJ was obtained at 275 nm (average power 1.03 W).**

Ultraviolet (UV) laser pulses are in great demand for a wide range of applications, including photolithography [1], spectroscopy [2] and femtosecond pump-probe measurements [3]. Despite this, the range of available sources is limited. Apart from large-scale synchrotrons and free-electron lasers, very few lasers directly emit UV light, examples being excimer and some solid-state lasers (e.g., cerium-based [4]). An alternative approach involves upconversion of visible or near-infrared laser light via the generation of discrete harmonics in $\chi^{(2)}$ and $\chi^{(3)}$ media [5,6]. Using an optical parametric amplifier as pump laser provides wavelength-tunability in the UV, and with careful design very short pulse durations can be achieved [7]. Although such systems are flexible, they are also complex, and repetition rate scaling is challenging because of the high pump energies required.

On the other hand, fiber and thin-disk technology allows repetition rate scaling in the near-infrared, pushing the frontiers of ultrafast lasers to unprecedented average power levels [8,9]. Much research is devoted to using these lasers for the generation of extreme UV light via high-harmonic generation [10,11]. To this end, pulse compression schemes are commonly employed, based for example on spectral broadening in bulk material [12] or gas-filled capillary and hollow-core photonic crystal fibers (HC-PCFs). This has allowed pulse compression from hundreds of fs to the few-cycle regime in set-ups involving two fiber stages [13,14]. In capillary fibers, spectral broadening normally occurs via self-phase modulation (SPM) in the normal dispersion regime, which requires the use of negatively chirped mirrors after the fiber and compresses the pulses in time by compensating for the induced chirp. If instead soliton-effect self-compression in the anomalous dispersion regime is used, there is no need for dispersion compensation. While this is difficult to achieve in large-bore capillary fibers (anomalous dispersion can only be achieved at very low gas pressures), it is straightforward in HC-PCFs, which deliver low loss even for small core diameters and which are ideal for soliton self-compression at energies in the µJ to tens of µJ range [15]. This removes the bandwidth restrictions imposed by chirped mirror technology, allowing generation of single-cycle pulses in a simple set-up [16].

A distinguishing feature of soliton self-compression in the presence of higher-order effects is dispersive wave (DW) generation, which is the result of phase-matching between the compressed soliton and linear waves [17,18]. DW emission in the deep and vacuum UV has been demonstrated in gas-filled HC-PCFs [19–21], and deep UV pulses with 72 nJ energy have been generated at 9.6 MHz repetition rate [22]. High-repetition rate (38 MHz) dispersive wave emission in the visible, with 13 nJ pulse energy, has also been reported [14].

Here we report the use of DW emission from self-compressed solitons in gas-filled HC-PCF to generate µJ pulses in the deep UV at close to 2 MHz repetition rate. The pump source is a compact commercial ytterbium fiber laser (Active Fiber Systems GmbH) that delivers ~300-fs-long pulses at 1030 nm with pulse energies up to tens of µJ. Because these pulses are too long for direct generation of coherent UV pulses via DW emission in gas-filled HC-PCF [23], it was necessary to compress them in a first HC-PCF stage before launching them into a second HC-PCF stage for UV generation.

The experimental set-up is sketched in Fig. 1(a). A single-ring PCF (SR-PCF - Fig. 1(c)) consisting of seven thin-walled capillaries arranged around a hollow core, was used. Guiding by anti-resonant reflection, SR-PCF has gained popularity because of its excellent guidance properties and simple structure. It provides broadband transmission bands with low loss (as low as 7.7 dB/km [24]), although phase-matched coupling to core-wall resonances creates high-loss bands at wavelengths $\lambda_q = [2t/(1+q)](n^2-1)^{1/2}$, where $t$ is the capillary wall thickness, $q$ the order of the resonance and $n$ is the refractive index of the glass [25]. Carefully designed fibers with small values of $t$ provide good transmission in the near-infrared and visible spectral regions, and the impact of the loss bands on the pulse dynamics can be minimized by shifting them far from the pump wavelength. Away from these bands the modal refractive index can be approximated by a capillary model [25], yielding anomalous dispersion in the evacuated fiber. When the fiber is filled with a noble gas, the optical Kerr effect and gas ionization are the dominant nonlinearities (the light-glass overlap is very small, and the system is effectively Raman-free [23]).

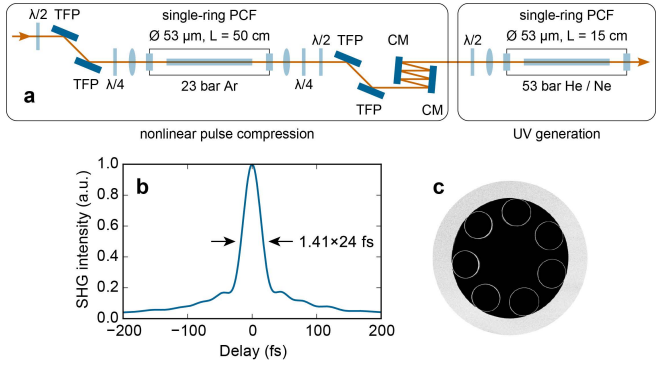

**Fig. 1.** (a) Experimental set-up. The fibers are placed in gas cells which can handle pressures up to 60 bar. TFP: thin-film polarizer, CM: negatively chirped mirror. Half-wave plates and TFPs are used for power control. (b) Autocorrelation measurement of the compressed pulses with 17.4 µJ energy at 1.92 MHz repetition rate. Assuming a Gaussian pulse shape, a pulse duration of 24 fs can be inferred from the autocorrelation. The long pedestal is due to uncompensated nonlinear phase from the pump laser. (c) Scanning electron micrograph of the single-ring PCF with 53 µm core diameter. The same fiber is used in both stages, however in the second stage it was HF-etched to reduce the wall thickness of the capillaries from initially ~350 to 220 nm.

In the nonlinear compression stage, the pulses from the laser were first broadened via SPM in a noble-gas-filled SR-PCF and then compressed with negatively chirped mirrors. Raising the gas pressure increases the nonlinearity, which reduces the energy required for pulse compression. This provides a means of adjusting the pulse energy of the compressor and thus making the most effective use of the available laser power. In the experiments the compressor was designed to operate at an input energy of ~20 µJ, the spectral broadening being achieved in a 50-cm-long SR-PCF with 53 µm core diameter and ~350 nm capillary wall thickness (Fig. 1(c)), filled with 23 bar of Ar. The input pulses were circularly polarized, which reduces the effective Kerr nonlinearity to 2/3 of its value for linear polarization. In this way a higher gas pressure could be used for a given input energy, resulting in weaker anomalous dispersion and higher quality compressed pulses. After the fiber, the polarization was converted back to linear. The negatively chirped mirrors (UltraFast Innovations GmbH) introduced a group delay dispersion of −2100 fs² to compensate for the positive chirp induced by SPM in the fiber and the dispersion of the optics up to the input of the second fiber. The pulse duration was measured by tapping off the pulses (after the nonlinear compression stage, Fig. 1(a)) and delivering them to a commercial autocorrelator through two 12.7-mm-thick fused silica plates that mimicked the dispersion of the optical elements used to launch light into the second fiber. At repetition rates from 100 kHz to 1.92 MHz, ~20 µJ pulses with duration >300 fs (between 311 and 331 fs, depending on the repetition rate) were compressed to ~25 fs (full-width-half-maximum - FWHM) with more than 82% transmission through the compressor (including all optics). At the highest repetition rate, 329 fs pulses with 20.5 µJ energy (39.3 W average power) were compressed to 24 fs, 17.4 µJ (33.5 W) with 85% transmission (Fig. 1(b)).

In the second fiber stage, we used the same SR-PCF, HF-etched to reduce the capillary wall thickness to ~220 nm. This shifted the $q=0$ loss band from ~740 to 470 nm, reducing its impact on the pulse dynamics. The fiber could in principle be etched further to allow loss-band-free transmission from the deep UV to the near-infrared. The output of the second fiber was collimated with an aluminium-coated off-axis parabolic mirror (Newport Corporation) and sent through a magnesium fluoride prism (Korth Kristalle GmbH) to spatially separate the UV. The power in the UV was then measured with a thermal power meter and corrected for the transmission of the gas cell window (5-mm-thick magnesium fluoride, Thorlabs, Inc.) and the reflectivity of the parabolic mirror. The prism was cut for Brewster angle transmission at 200 nm, resulting in small Fresnel losses at the prism surfaces over the entire UV spectral region. A fiber-coupled CCD spectrometer was used to measure the spectrum.

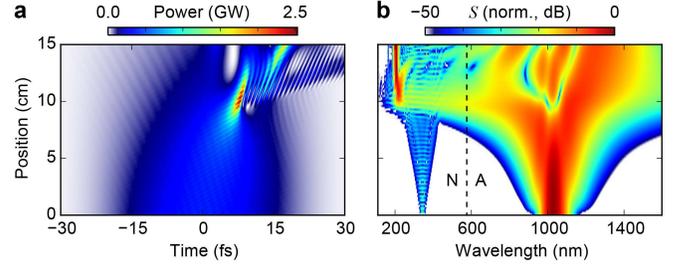

**Fig. 2.** Simulated temporal (a) and spectral (b) evolution of a 25-fs-long Gaussian pulse with 10 µJ at 1030 nm in a SR-PCF with 53 µm core diameter filled with 53 bar He. The spectral energy density $S$ is shown per unit wavelength, normalized to its peak value. The zero-dispersion wavelength (dashed line) is at 576 nm. N and A denote regions of normal and anomalous dispersion. The input pulse self-compresses to a duration of 1.2 fs FWHM before emitting a DW at 205 nm. The DW is emitted with a duration of 10 fs and broadens to 15 fs at the fiber output. In practice, this temporal broadening can be avoided by optimizing the fiber length.

To investigate pulse propagation in the second fiber, we performed numerical simulations based on a unidirectional field equation [26]. An example is shown in Fig. 2. The pump pulses undergo soliton dynamics when the gas pressure is adjusted for anomalous dispersion. At sufficiently high input energies, they initially compress temporally and broaden spectrally, as expected of higher-order solitons. In the experiments, the soliton order was below ~6, resulting in coherent dynamics [27]. As the pulses compress, higher-order effects come into play. Self-steepening and the optical shock effect at the trailing edge of the pulses lead to strongly asymmetric spectral broadening towards shorter wavelengths. Phase-matching to DWs in the deep UV is made possible by higher-order dispersion, and occurs when $\beta(\omega_{DW}) = \beta_{sol}(\omega_{DW})$, where $\omega_{DW}$ is the DW frequency and $\beta$ and $\beta_{sol}$ are the propagation constants of linear waves and solitons. When this condition is not exactly satisfied the dephasing rate takes the form:

$$\vartheta(\omega) = \beta(\omega) - \left\{ \beta_0 + \beta_1 [\omega - \omega_{sol}] + \gamma P_P \frac{\omega}{\omega_{sol}} \right\}, \quad (1)$$

where $\beta_0$ is the propagation constant and $\beta_1$ the inverse group velocity at the soliton center frequency $\omega_{sol}$, $\gamma$ is the gas-dependent nonlinear fiber parameter at $\omega_{sol}$ and $P_P$ is the soliton peak power. Fig. 3(a) plots $\vartheta(\omega)$ versus wavelength for a given set of fiber and gas parameters. The DW wavelength is widely tunable via core diameter, gas species and pressure. Increasing the gas pressure or using a heavier gas red-shifts the phase-matching point, while an increase of the peak power $P_P$ leads to a blue-shift. Additionally, the pump pulses can undergo substantial blue-shifting (through ionization) before DW emission, leading to a DW red-shift. The use of lighter gases, which have higher ionization potentials, typically improves the conversion efficiency to the UV through reduced ionization and cleaner self-compression. We therefore chose the lightest gas that allowed phase-matching at the target wavelength (up to the maximum pressure of the gas cell – 60 bar).

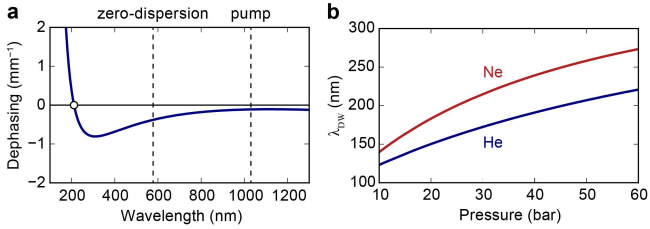

**Fig. 3.** Phase-matching to DWs according to Eq. (1). The fiber has a core diameter of 53 µm. The pump wavelength is 1030 nm, and the peak power is 1 GW. (a) Fiber filled with 53 bar He. Phase-matching occurs at 211 nm (circle). The shallow dispersion of the gas-filled fiber yields long coherence lengths for broadband DW emission (here the coherence length exceeds 3 cm for a bandwidth of 15 nm/103 THz, corresponding to a relative bandwidth of 7.3%). (b) Phase-match wavelength $\lambda_{DW}$ as function of pressure for He and Ne.

We filled the second fiber with He and Ne to demonstrate DW emission in the deep UV at ~200 and ~270 nm, respectively. Fig. 4(a) shows the measured spectrum after the second fiber when it was filled with 53 bar He and pumped with 17 µJ pulses at 100 kHz repetition rate. In this case, a DW was emitted at 205 nm with 1.05 µJ energy (6.2% of the input to the fiber), corresponding to an average power of 105 mW. The UV beam profile was in a clean fundamental mode, and the DW spectrum was broadband (7.9 nm/56 THz FWHM), supporting transform-limited pulses of 3.8 fs duration. It has previously been shown that DWs emitted from self-compressed solitons in gas-filled HC-PCFs are ultrashort (sub-5-fs in the work of Ermolov et al. [28]), an observation that is confirmed by the numerical simulations presented here. For DW emission at longer wavelengths, the fiber was filled with 53 bar Ne. In this case, upon pumping with 9 µJ pulses at 100 kHz repetition rate, a DW was emitted at 264 nm, with 0.74 µJ energy (8.2% of the input to the fiber), corresponding to an average power of 74 mW.

Having demonstrated high-energy DW emission in the deep UV, our next aim was to study how the process scales with repetition rate. Modest pump pulse energies of 10–20 µJ allow MHz repetition rates (up to 1.92 MHz in the current experiments) at only tens of watts average power. The results are summarized in Fig. 5. While the average power in the UV increased with repetition rate, the pulse energy decreased. Nevertheless, an average power of 1.03 W could be obtained at 275 nm (fiber filled with Ne, 1.92 MHz repetition rate). The corresponding spectrum is shown in Fig. 4(b). In this case, the pulse energy in the UV was still 0.54 µJ, which means that 6% of the input energy to the fiber was converted to the UV.

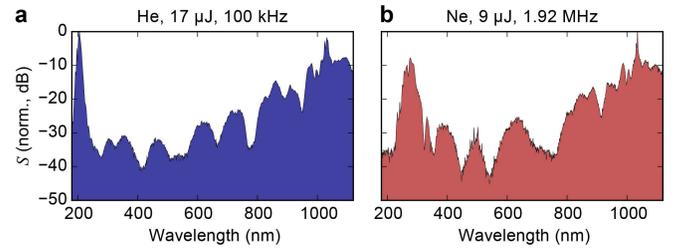

**Fig. 4.** (a) Measured spectrum from the second fiber when filled with He and pumped with 17 µJ pulses at 100 kHz repetition rate. The pulse energy in the UV was 1.05 µJ (105 mW average power). The spectrum is not intensity-calibrated below 200 nm, where the sensitivity of the spectrometer decreases rapidly. (b) Spectrum for fiber filled with Ne and pumped with 9 µJ pulses at 1.92 MHz. In this case, the pulse energy in the UV was 0.54 µJ (1.03 W average power).

As the pump energy is increased, the DW emission band typically shifts towards shorter wavelength. This is because the self-compressed pulses have higher peak power, strengthening the Kerr contribution to the dephasing parameter (Eq. (1)). This behavior was observed for all repetition rates in the Ne-filled fiber, but only at the lowest repetition rate (100 kHz) in the He-filled fiber. Instead, at 505 kHz and 1.01 MHz, the DWs initially blue-shifted with increasing input energy, but then red-shifted as the input energy was further increased. At the highest repetition rate (1.92 MHz) the DWs continuously shifted towards longer wavelength with increasing input energy, reaching 273 nm at the highest input energy, i.e., 68 nm/363 THz from the emission wavelength for 100 kHz repetition rate (Fig. 5(c)).

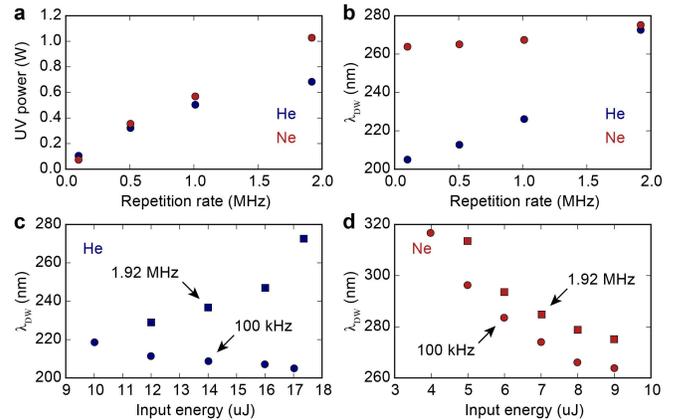

**Fig. 5.** Power in the UV (a) and wavelength of the DWs (b) as function of the repetition rate when the fiber was filled with He (blue) and Ne (red). Wavelength of the DWs as function of the input energy for the He-filled (c) and Ne-filled (d) fiber, for 100 kHz (dots) and 1.92 MHz (squares) repetition rate.

A much smaller repetition-rate dependence of the DW wavelength was observed in the Ne-filled fiber, with a red-shift of only 11 nm/47 THz (264 to 275 nm) when the repetition rate was increased from 100 kHz to 1.92 MHz (Fig. 5(d)). Although this red-shift with increasing repetition rate can be partially compensated for by decreasing the gas pressure, this is at the expense of reduced conversion efficiency to the UV.

We attribute these DW red-shifts to slow gas dynamics inside the fiber, caused for example by heating through nonlinear absorption and ionization. In free-space experiments it has been shown that similar dynamics evolve on timescales from ns to ms [29], so we expect that this also occurs in fiber, especially at high repetition rates when the system does not fully recover between successive laser shots. An even stronger difference between low (100 kHz) and high (1.92 MHz) repetition rates is evident when the DW emission is at shorter wavelength (~205 nm, Fig. 5(c)). This is because the strong shock effect required for efficient DW generation at ~205 nm causes stronger ionization of the gas. If, however, the DW emission is tuned to longer wavelength the required pump pulse compression is less extreme, reducing the ionization level and allowing better scaling with repetition rate (Fig. 5(d)). Since the dynamics at high repetition rates are clearly interesting for many systems operating in the strong-field regime at MHz repetition rates, we plan to investigate them in future experiments.

In conclusion, high-energy deep UV pulses can be generated in gas-filled HC-PCF at repetition rates from 100 kHz to 1.92 MHz via DW emission. At 100 kHz, pulse energies >1 µJ were obtained at 205 nm, while at 1.92 MHz, more than 1 W average power was obtained at 275 nm. Although no degradation of the SR-PCF was observed during the experiments, which lasted several hours, further studies will be required to assess the long-term stability of the system. Together with a pump laser that is carrier-envelope-phase stable, scaling to even higher repetition rates suggests the exciting possibility of bright frequency combs in the deep and maybe even the vacuum UV. Finally, the emission of DWs at short wavelengths, as demonstrated in this work, requires strong temporal self-compression of the input pulses to the single-cycle regime. Such short pulses, with peak powers exceeding the GW-level that can be directly delivered to a gas jet for high-harmonic generation [30], could in the near future be used to generate isolated attosecond pulses at MHz repetition rates.

**Acknowledgments**. We thank M. H. Frosz for fabricating the fiber used, and C. Martens Biersach and R. Keding for HF-etching.